\pdfoutput=1 

\documentclass{article}
\usepackage{arxiv}

\usepackage{url,hyperref,lineno,microtype}
\usepackage[onehalfspacing]{setspace}

\usepackage{amsmath, bm}
\usepackage{amsfonts}
\usepackage{mathtools}
\usepackage{siunitx}

\usepackage{mathabx}
\usepackage{booktabs}

\usepackage{subcaption}
\usepackage{graphicx}
\usepackage[dvipsnames,table]{xcolor}

\usepackage{hyperref,url}
\usepackage[english]{babel}
\newtheorem{ex}{Example}
\usepackage{changes}
\newtheorem{rem}{Remark}
\newtheorem{defi}{Definition}
\usepackage[utf8]{inputenc} 
\usepackage[T1]{fontenc}    
\usepackage{nicefrac}       
\usepackage{cleveref}       
\usepackage{natbib,hyperref}
\usepackage{doi}

\title{Partial Directed Coherence and the Vector Autoregressive Modelling Myth and a Caveat}

\author{ \href{https://orcid.org/0000-0002-2668-6747}{\includegraphics[scale=0.06]{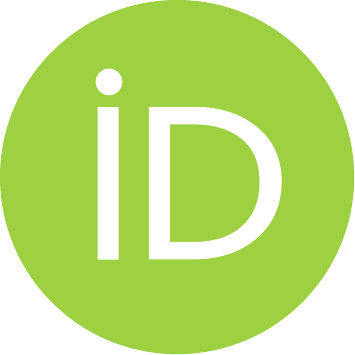}\hspace{1mm}Luiz A.~Baccalá}\thanks{Corresponding author.} \\
	Laborat\'{o}rio de Comunica\c{c}\~{o}es e Sinais\\
	Departamento de Telecomunicações e Controle\\
	Escola Polit\'{e}cnica\\
	Universidade de S\~{a}o Paulo\\
	S\~{a}o Paulo, Brazil\\
	\texttt{baccala@lcs.poli.usp.br} \\
	\And
	\href{https://orcid.org/0000-0002-0290-2927}{\includegraphics[scale=0.06]{orcid.pdf}\hspace{1mm}Koichi Sameshima} \\
	Laborat\'{o}rio de Investiga\c{c}\~{a}o M\'{e}dica 43 -- HCFMUSP\\
	Departamento de Radiologia e Oncologia\\
	Faculdade de Medicina\\
	Universidade de S\~{a}o Paulo\\
	S\~{a}o Paulo, Brazil \\
	\texttt{ksameshi@usp.br} \\
}

\renewcommand{\shorttitle}{PDC: ditching the VAR Myth}

\begin{document}
\maketitle
\begin{abstract}
Here we dispel the lingering myth that Partial Directed Coherence is a Vector Autoregressive (VAR) Modelling dependent concept. In fact, our examples show that it  is \textit{spectral factorization}  that lies at its heart, for which VAR modelling is a mere, albeit very efficient and convenient, device. This applies to Granger Causality estimation procedures in general and also includes instantaneous Granger effects. Care, however, must be exercised for connectivity between multivariate data generated through nonminimum phase mechanisms as it may possibly be \textit{incorrectly} captured.
\end{abstract}


\section{Introduction}

The aim of Granger time series connectivity modelling is to examine how observations from different simultaneously observed time series may be related in the hope of exposing possible mechanisms behind their 
generation. This goal is intrinsically limited by a number of factors: chief among them are potential structural artifacts that result from unobserved series (confounders). This plus the fact that Granger analysis rests exclusively on observations rather than active intervention ~\citep{baccala_causality_2014} means that one must characterize interactions as `Granger-causal' rather than causal in the strictest sense.

In spite of this, and in connection to situations where intervention is either impossible, such as when impacting phenomena on a geophysical scale as for Solar spot/Melanoma data ~\citep{Baccala-2014Pdc} or undesirable as in physiological data analysis where noninvasive methods, at least in the human case, are always to be preferred,
Granger Causality remains of interest in providing clues as to the dynamics behind the observed variables.

In recent years a vast array of methods have been developed; they originated in economics research following Granger's 1969 seminal paper \citep{Granger1969} who used vector autoregression as a device to model data relationships in the time domain. His `causality' notion rests on how well the knowledge of one time series's  past can enhance one's ability to predict another time series, which once vindicated, implies their connectivity. Though initially a strictly bivariate concept, the idea has been extended to the analysis of more than two simultaneously observed time series in an attempt to disentangle the effect of other series that might be acting as interaction confounders to pairwise observations ~\citep{Pbrovercoming}. Historically, most developments that followed rest on \citet{Geweke1984}'s work who used Vector Autoregressive (VAR) modelling for more than two time series as a preliminary step to deduct the effect of the other observed series from the time series pair of interest. After that subtraction, the method consists of looking at a power ratio of the prediction errors between when the past of a series is taken into account against when it is not. 

Much along the lines of improved estimation and inference of Granger time domain representations has been made since and can be read in ~\citep{Lutkepohl2005}.

As a general rule, much of what followed is patterned on the representation of temporal data in terms of  `output-only' systems, i.e. systems where the observed time series, $x_1(n),\dots, x_N(n)$, are represented as conveniently filtered versions of white noise --- the so called innovations. 

Because VAR models can be naturally  interpreted in terms of linear filtering, already some aspects of a spectral interpretation to the Granger connectivity scenario were present in \citep{Geweke1984}'s work. Further specifics have been developed since~\citep{Lutkepohl2005,Barrett2010}. 

The spectral nature of these problems, specially in connection to EEG data processing which are naturally characterized in terms of oscillatory behaviour, was boosted by the introduction of \textit{Directed Transfer
Function} (DTF)  ~\citep{Kaminski1991} and later by \textit{partial directed coherence} (PDC) \citep{Baccalabiocib2001}. Both quantities employed VAR modelling for their definition. Also both have since evolved to more accurate, and thus, more appropriate measures, please see ~\citep{Baccala2021} for their development. A \textit{leitmotif} of those improvements was the growing realization of the importance and consequent incorporation of the estimated covariance of the innovations noise driving the observed outputs $x_i(n)$ ~\citep{gPDC2007,Takahashi2010,Baccala2021,entropy2021}. 

In fact, explicit consideration of innovations covariance effects are important in connection to the so-called `instantaneous' Granger causality (iGC) and are helpful in unveiling aspects of cardio-hemodynamic behaviour ~\citep{faesLivro}. Much as in the case of GC itself, iGC was originally only seen as time domain aspect. There have been early efforts to portray it in the frequency domain ~\citep{Luc-2010b,faesLivro}; more general efforts have only recently appeared with  ~\citet{Cohen-2019} and \citet{Nuzzi2021} along Geweke's line of description and along PDC/DTF lines ~\citep{entropy2021}.

All of the latter developments have relied heavily on VAR modelling. This paper, by contrast, aims to dispel the notion that  PDC \citep{Baccalabiocib2001} (and DTF, its dual)  or any of its related quantifiers require vector autoregressive (VAR) modelling as a mandatory prerequisite. 
This notion coupled with limited familiarity with VAR modelling may have been a hindrance to their spread as methods of choice for Granger time series connectivity modelling among non time series specialists. We show here that absolute reliance on VAR modelling is not a must, but rather a matter of convenience, even though PDC and DTF were originally introduced with the help of VAR models. 

As we have been alerted in the review process to this paper, an early precursor to the present developments is contained in ~\citep{Jachan2009}, which undeservedly does not seem to have attracted much following having  just 22 citations at the Web of Science at the moment of this writing with only a small fraction of them reflecting actual practical method employment, mostly by its proponents. The present exposition  not only confirms those results but provides evidence that they hold for more general PDC/DTF versions as well.

To dispel the VAR reliance misconception we employ a set of examples comprising a variety of methods, parametric and nonparametric, that, as we show next, yield essentially the same results. The methodological equivalence between them holds even for total PDC (tPDC)  and total DTF (tDFT)  as defined in \citep{entropy2021} which represent recently introduced extensions that incorporate the effects of instantaneous Granger causality
 to connectivity descriptions.
 
For brevity, we only show results for total PDC since it incorporates a consistent frequency domain description of instantaneous Granger interactions to PDC that automatically extends to total DTF's, given their  duality \citep{Baccala2021,entropy2021}. 

The rest of this paper is organized as follows: Sec. \ref{sec:mathcon} reviews the theoretical basis and is followed in Sec. \ref{sec:estim} with a brief description of the methods employed in the comparative computations which are illustrated in Sec. \ref{sec:numillum} and commented in Sec. \ref{sec:discuss} leading to the conclusion in Sec. \ref{sec:conclu} that tPDC/PDC (tDTF/DTF) representations are essentially canonical factors of the joint power spectral density of the data which portrays the relationship between multivariate data.

A concept that turns out to be key in the present setup is that of \textit{spectral factorization} and the notion of a \textit{minimum phase} spectral factor covered in more detail in Sec. \ref{sec:linmodels}.

The concept of a \textit{minimum} versus a \textit{nonminimum} phase system is important for our discussion. This is briefly examined in the development that follows as we show it can lead to possibly false connectivity inference when they nonminimum phase mechanisms are behind the data generation process.


\section{Mathematical Considerations}
\label{sec:mathcon}
\subsection{General Linear Models with Rational Spectra}
\label{sec:linmodels}
A general class of linear stationary multivariate processes $\mathbf{x}(n)=[x_1(n) \dots x_N(n)]^T$ is represented \citep{Lutkepohl2005} by:

\begin{equation}
\label{eq:linmodel}
\mathbf{x}(n)=\sum_{r=1}^p \boldsymbol{\mathfrak{A}}_r \boldsymbol{x}(n-r) +\sum_{s=0}^q \boldsymbol{\mathfrak{B}}_s \mathbf{w}(n-s),
\end{equation}

\noindent where $\mathbf{w}(n)=[w_1(n) \dots w_N(n)]^T$ is a stationary (zero mean without loss of generality) multivariate innovations process with covariance matrix $\boldsymbol{\Sigma}_{\mathbf{w}}$. The process defined by \eqref{eq:linmodel} is termed a Vector Autoregressive Moving Average process, denoted VARMA($p$,$\,q$),  whose structure is defined by the $\boldsymbol{\mathfrak{A}}_r, \boldsymbol{\mathfrak{B}}_s$ matrices \citep{Lutkepohl2005}. VAR processes and vector moving average (VMA) processes are special cases, when  or $\boldsymbol{\mathfrak{B}}_s=0,\forall  s>0$, or $\boldsymbol{\mathfrak{A}}_r=0, \forall r$, respectively.  The equivalences between VAR($p$) and VMA($\infty$), and between VMA($q$) and VAR($\infty$) are well known, where $p$ and $q$ refer respectively to the AR and MA orders that make up the model. 

We implicitly assume that \eqref{eq:linmodel} is stable, i.e. the associated $\mathbf{x}(n)$ is wide sense stationary. For simplicity we consider only the case of finite $p$ and $q$. This is guaranteed if the magnitude of the roots of
\begin{equation}
\label{eq:Aroots}
det \; \boldsymbol{\mathfrak{A}}(z)=0
\end{equation}
are less than 1 for
\begin{equation}
\label{eq:arpartz}
\boldsymbol{\mathfrak{A}}(z)=\mathbf{I}-\sum_{r=1}^p\boldsymbol{\mathfrak{A}}_r z^{-r}
\end{equation}
where $det$ stands for the determinant.
\begin{defi}
\label{def:minimumphase}
The system represented by \eqref{eq:linmodel} is \textit{minimum phase}  if the magnitude of the roots of
\begin{equation}
\label{eq:detB}
det \; \boldsymbol{\mathfrak{B}}(z)=0
\end{equation}
are less than or equal to 1 for
\begin{equation}
\label{eq:mapartz}
\boldsymbol{\mathfrak{B}}(z)=\sum_{s=0}^q\boldsymbol{\mathfrak{B}}_s z^{-s}
\end{equation}
\end{defi} 

Definition \ref{def:minimumphase}  guarantees  that stable $\mathbf{w(n)}$ innovations sequences for $n \geq 0$ may be found that lead to the observations, i.e. the system  defined by \eqref{eq:linmodel} has a stable inverse.
\begin{rem}
Strictly speaking when the roots in \eqref{eq:mapartz} are equal to 1, the impulse response of the inverse is merely bounded.
\end{rem}

\begin{rem}
When used as a  data generating mechanism for $\mathbf{x}(n)$ data, \eqref{eq:linmodel} 
does not need to be minimum phase. However, data modelling through \eqref{eq:linmodel} always leads to an estimated minimum phase counterpart system. This follows from the fact that only second order statistics are used for estimating \eqref{eq:linmodel} coefficients. When the data is Gaussian, this is the only available alternative, as higher order statistics are redundant and offer no additional information that might expose any evidence of possible phase nonminimality.
\end{rem}

It is easy to show that the power spectral density matrix of $\mathbf{x}(n)$ \eqref{eq:linmodel} is given by:
%

\begin{equation}
\label{eq:powerspec}
\mathbf{S}_\mathbf{x}(\nu)=\boldsymbol{\mathfrak{A}}^{-1}\!(\nu)\,\boldsymbol{\mathfrak{B}}(\nu)\,\boldsymbol{\Sigma}_{\mathbf{w}}\,\boldsymbol{\mathfrak{B}}^H\!(\nu)\,\boldsymbol{\mathfrak{A}}^{-H}\!(\nu),
\end{equation}
where
\begin{equation}
\label{eq:arpart}
\boldsymbol{\mathfrak{A}}(\nu)=\mathbf{I}-\sum_{r=1}^p\boldsymbol{\mathfrak{A}}_r e^{-\mathbf{j}2\pi r \nu}
\end{equation}

\begin{equation}
\label{eq:mapart}
\boldsymbol{\mathfrak{B}}(\nu)=\sum_{s=0}^q\boldsymbol{\mathfrak{B}}_s e^{-\mathbf{j}2\pi s \nu},
\end{equation}

\noindent for $0\leq |\nu|<0.5$ which represents the normalized frequency and $\mathbf{j}=\sqrt{-1}$. Naturally \eqref{eq:arpart} and \eqref{eq:mapart} are associated with making $z=e^{\,\mathbf{j}2\pi r \nu}$ in \eqref{eq:arpartz} and \eqref{eq:mapartz} respectively.

It is easy to realize that \eqref{eq:powerspec} is of the form 


\begin{equation}
\label{eq:powerform}
\mathbf{S}_\mathbf{x}(\nu)=\boldsymbol{\mathfrak{H}}(\nu)\;\!\boldsymbol{\Sigma}_{\mathbf{w}}\;\!\boldsymbol{\mathfrak{H}}^H\!(\nu)
\end{equation}
containing the frequency dependent factor, $\boldsymbol{\mathfrak{H}}(\nu)$, and a frequency independent factor, $\boldsymbol{\Sigma}_{\mathbf{w}}$.
\begin{rem}
Equations \eqref{eq:powerspec} and \eqref{eq:powerform} hold regardless of whether \eqref{eq:linmodel} is minimum phase or not.
\end{rem}

From \eqref{eq:powerform} it is easy to write the coherency matrix $\boldsymbol{\mathfrak{C}}(\nu)$ with entries:
\begin{equation}
\label{eq:cohere_def}
C_{ij}(\nu)=\dfrac{S_{ij}(\nu)}
{\sqrt{S_{ii}(\nu) \; S_{jj}(\nu)}}
\end{equation}
 by writing


\begin{eqnarray}
\boldsymbol{\mathfrak{C}}(\nu)&=&\mathfrak{D}(\mathbf{S}_\mathbf{x}(\nu))^{-1/2}\,\mathbf{S}_\mathbf{x}(\nu)\,\mathfrak{D}(\mathbf{S}_\mathbf{x}(\nu))^{-1/2}\nonumber\\
&=&\mathfrak{D}(\mathbf{S}_\mathbf{x}(\nu))^{-1/2}\,\boldsymbol{\mathfrak{H}}(\nu)\,\boldsymbol{\Sigma}_{\mathbf{w}}\,\boldsymbol{\mathfrak{H}}^H\!(\nu)\,\mathfrak{D}(\mathbf{S}_\mathbf{x}(\nu))^{-1/2}\nonumber\\
&=&\boldsymbol{\Gamma}(\nu)\,\boldsymbol{\mathfrak{R}}\,\boldsymbol{\Gamma}^H\!\:\!(\nu)\label{eq:cohere_Matrix}
\end{eqnarray}

\noindent where $\mathfrak{D}(\cdot)$ is the diag matrix operator, i.e. one that produces a matrix that is nonzero except for preserving the diagonal elements of the operand so that


\begin{equation}
\label{eq:Gamma}
\boldsymbol{\Gamma}(\nu)=\mathfrak{D}(\mathbf{S}_\mathbf{x}(\nu))^{-1/2}\boldsymbol{\mathfrak{H}}(\nu)\,\mathbf{D}^{1/2}
\end{equation}

\noindent and


\begin{equation}
\label{eq:corrmatrix}
\boldsymbol{\mathfrak{R}}=\mathbf{D}^{-1/2}\boldsymbol{\Sigma}_{\mathbf{w}}\mathbf{D}^{-1/2}
\end{equation}

\noindent is a correlation matrix with ones along the main diagonal for $\mathbf{D}=\mathfrak{D}(\boldsymbol{\Sigma}_{\mathbf{w}})$. 

Writing \eqref{eq:cohere_Matrix} as a product of the frequency dependent part $\boldsymbol{\Gamma}(\nu)$ mediated by a correlation matrix $\boldsymbol{\mathfrak{R}}$ allows one to apply the definition of 
total DTF matrix \citep{entropy2021} as:


\begin{equation}
\label{eq:tDTF}
\overgroup{\boldsymbol{\Gamma}}\!\!(\nu)=\boldsymbol{\Gamma}(\nu)\odot \boldsymbol{\Gamma}^*\!(\nu)+\boldsymbol{\Gamma}(\nu)\boldsymbol{\rho} \odot\boldsymbol{\Gamma}^*\!(\nu)
\end{equation}

\noindent where $\boldsymbol{\rho}=\boldsymbol{\mathfrak{R}}-\mathbf{I}_N$, and $\mathbf{I}_N$ is an $N\times N$ identity matrix with $\odot$ standing for the Hadamard element-wise matrix product.

The entries $i,j$ from $\overgroup{\boldsymbol{\Gamma}}\!\!\:(\nu)$ reduce to the absolute square value of directed coherence from $j$ to $i$, which is a scale invariant form of DTF \citep{Baccala1998a}, when instantaneous Granger causality is absent. Equation \eqref{eq:tDTF} describes what we have termed \textit{Total Granger Influentiability} ~\citep{entropy2021}. 

An entirely parallel development allows defining total partial directed coherence \citep{entropy2021}, taking advantage of the fact the partial coherence matrix can be shown to equal:


\begin{eqnarray}
\label{eq:partial_matrix}
\boldsymbol{\mathfrak{K}}(\nu)&=&\boldsymbol{\mathfrak{C}}^{-1}(\nu)\nonumber\\
&=&\boldsymbol{\Pi}^H\!(\nu)\,\boldsymbol{\mathfrak{\widetilde{R}}}\,\boldsymbol{\Pi}(\nu)
\end{eqnarray}
for 
\begin{equation}
\label{eq:pi}
\boldsymbol{\Pi}(\nu)=\mathbf{D}^{1/2}\boldsymbol{\mathfrak{H}}^{-H}\!(\nu)\,\mathfrak{D}(\mathbf{S}_\mathbf{x}(\nu))^{1/2}\,\mathbf{\tilde{D}}^{1/2}
\end{equation}
and 
\begin{equation}
\boldsymbol{\mathfrak{\widetilde{R}}}=\mathbf{\tilde{D}}^{-1/2}\,\boldsymbol{\Sigma}^{-1}_{\mathbf{w}}\,\mathbf{\tilde{D}}^{-1/2}
\end{equation}

\noindent which is a partial correlation matrix between the $w_i(n)$ innovations where $\mathbf{\tilde{D}}=
\mathfrak{D}(\boldsymbol{\Sigma}^{-1}_{\mathbf{w}})$.

The form in \eqref{eq:partial_matrix} is what allowed us to define total PDC as:
\begin{equation}
\label{eq:tPDC}
\overgroup{\boldsymbol{\Pi}}\!\!\:(\nu)=\boldsymbol{\Pi}^*\!(\nu)\odot \boldsymbol{\Pi}(\nu)+\boldsymbol{\Pi}^*\!(\nu)\odot\boldsymbol{\tilde{\rho}}\,\boldsymbol{\Pi}(\nu)
\end{equation}
where $\boldsymbol{\tilde{\rho}}=\mathbf{\mathfrak{\widetilde{R}}}-\mathbf{I}_N$. The $i,j$ entries  describe what we termed the \textit{Total Granger Connectivity} from $j$ to $i$ \citep{entropy2021}, which reduce to generalized PDC \citep{gPDC2007} when instantaneous Granger causality is absent.
 
Whenever one can properly write the spectral density matrix  as in \eqref{eq:powerform}, one may employ the latter quantities to describe multivariate time series within the tPDC$\,$-$\,$tDTF framework. A case in point which we describe briefly in Sec. \ref{sec:wilson} is provided by Wilson's spectral factorization algorithm \citep{Wilson-1972}, which has been used before in connection with alternative Granger causality characterizations \citep{Dhamala2008} and is also behind ~\citet{Jachan2009}'s results.

\section{Estimation Methods}
\label{sec:estim}
Equation \eqref{eq:linmodel} was used as a general data mechanism for imposing relationships between the time series we examine in Sec. \ref{sec:numillum}. The data generated were analysed via the three main approaches we briefly describe next.

\subsection{Vector Autoregressive Modelling}
\label{sec:varMod}

Vector autoregressive modelling is a traditional subject \citep{Lutkepohl2005}. The  version used here was implemented in the AsympPDC package \citep{asymp_pdc2} and employs Nuttall-Strand's method to obtain the autoregression coefficients \citep{marple_book}. One important step in this sort of procedure involves finding the best model order $p$. Here Hannan-Quinn's method  was chosen; it is a variant from the better known Akaike's method \citep{Lutkepohl2005}.

\subsection{Vector Moving Average and Vector Autoregressive Moving Average Modelling}
\label{sec:varmaMod}

A traditional means of fitting VMA($q$) and VARMA($p$,$\,q$) models is to determine a preliminary VAR model of very large order ($p=50$ was adopted here) and use its residuals $\epsilon_i(n)$ to fit the observed data  $x_j(n)$ through a mock multi-input/multi-output system via least-squares. An univariate version of this approach can be appreciated in \citep{StoicaSpectralanalysissignals2005}.

In practical applications, determining $p$ and $q$ can be achieved through minimizing model order choice functions as in Akaike's method. Whereas, minimizing Akaike-type penalization is trivial in the VMA case, bidimensional search of tentative $p$ and $q$  is required in the VARMA case. To simplify matters here, we have employed the theoretical model orders used to get the estimates.

\subsection{Wilson's Algorithm}
\label{sec:wilson}
 Wilson's method is an iterative method that decomposes \eqref{eq:powerform} into estimates for $\boldsymbol{\mathfrak{H}}(\nu)$ and $\boldsymbol{\Sigma}_\mathbf{w}$ \citep{Wilson-1972}. It starts by guessing a $\boldsymbol{\mathfrak{H}}(\nu)$ with the restriction of its representing filters to have impulse responses that are 
identically zero for negative time (the so-called filter causality condition, sometimes referred as nonanticipative filters whose output cannot anticipate the input). The solution essentially amounts to Newton's root finding iterations until a maximum prescribed error is achieved. In the present case, a maximum
error of $10^{-6}$ was adopted.

Wilson's method has been used before in connection with other Granger causality descriptions both related ~\citep{Jachan2009} and directly unrelated \citep{Dhamala2008} to PDC/DTF descriptions. It has the advantage that it can be applied to nonparametric spectral estimates, whether they are obtained by periodogram smoothing \citep{percival_book} or other means like wavelets \citep{LimaGrangercausalityfrequency2020}. 

The spectral estimates used here (henceforth referred as \textbf{WN}, nonparametric Wilson estimates) employed Welch's method as implemented in Matlab's \texttt{cpsd.m} function with von Hann's data window and 50\% segment overlap \citep{percival_book}. 



The reader may obtain a working Python implementation in \citep{LimaGrangercausalityfrequency2020}. Here a similar Matlab code version was used.

\subsection{Brief Comments}
\label{sec:comment_est}

The time series modelling methods of Sec. \ref{sec:varMod} and \ref{sec:varmaMod} are essentially least squares approaches. Wilson's algorithm on the other hand is a numerical square-rooting procedure that also achieves the spectral factorization of the power spectral density matrix $\mathbf{S}(\nu)$. In all cases, one obtains the so-called minimum phase spectral factor represented by $\boldsymbol{\mathfrak{H}}(\nu)$ in  \eqref{eq:powerform}. 

All Matlab routines used in this paper have been included as Supplementary Material. For convenience, Dhamala's most recent implementation \citep{Henderson-Dhamala2021} was also included and essentially leads to the same results we report next.

\section{Numerical Illustrations}
\label{sec:numillum}
In the following illustrations, the data comprise $n_s=16,384$ observed points to minimize misinterpretation due to short time series effects. In all cases, the theoretical models can be used to compute the theoretical total PDC as in \citep{entropy2021}. In each case, the mean-squared frequency domain approximation error of each estimation method was computed and is presented in Table \ref{tab:errors} after averaging over $R=100$ realizations. Here Wilson estimates employed $256$-point long data tappers.

Next we present three examples whose allied graphs contain the real and imaginary parts of tPDC plotted against the background of the expected theoretically computed results. These examples share the property of being generated by minimum phase \eqref{eq:linmodel} models. 

Finally, a fourth example generated by a nonminimum phase \eqref{eq:linmodel} is examined. Its numerical results are contrasted to the theoretical tPDC computed with help of the actual generating model parameters. 

\begin{ex} \textbf{Vector Moving Average Model (VMA)}
\label{ex:vma1}

We start with conceivably the simplest possible kind of vector moving average example with 
unidirectional influence and with the clear presence of iGC described by

\[ 
\left\{\begin{aligned}
x_1(n)&=& w_1(n)+w_2(n-1)\\ 
x_2(n)&=& w_2(n)+w_2(n-1)
\end{aligned} \right.
\label{eq:example1_vma} 
\]

\noindent with innovations noise covariance 
\begin{equation}
\label{eq:vma1_pf}
\boldsymbol{\Sigma}_{\mathbf{w}}=\left[
\begin{array}{cc}
1 & 1\\
1 & 5
\end{array}
\right]
\end{equation}
whose influence of $x_2(n)$ onto $x_1(n)$ is clear  due to its lagged dependence on $w_2(n)$ which is the sole input that determines $x_2(n)$. The presence of iGC is clear from \eqref{eq:vma1_pf}'s non diagonal nature.

From Figure ~\ref{fig:ex1}, it is clear that for large $n_s$, all estimates of total PDC agree with the theoretically expected one within the constraints of estimator nature. A case in point is Wilson's factorized version computed from the nonparametric power spectral estimates which is rippled as expected \added{(\textbf{\textcolor{red}{red lines}})}, following what happens with the original spectral estimates.

\begin{figure}[h!]
\begin{center}
\includegraphics[width=0.95\textwidth]{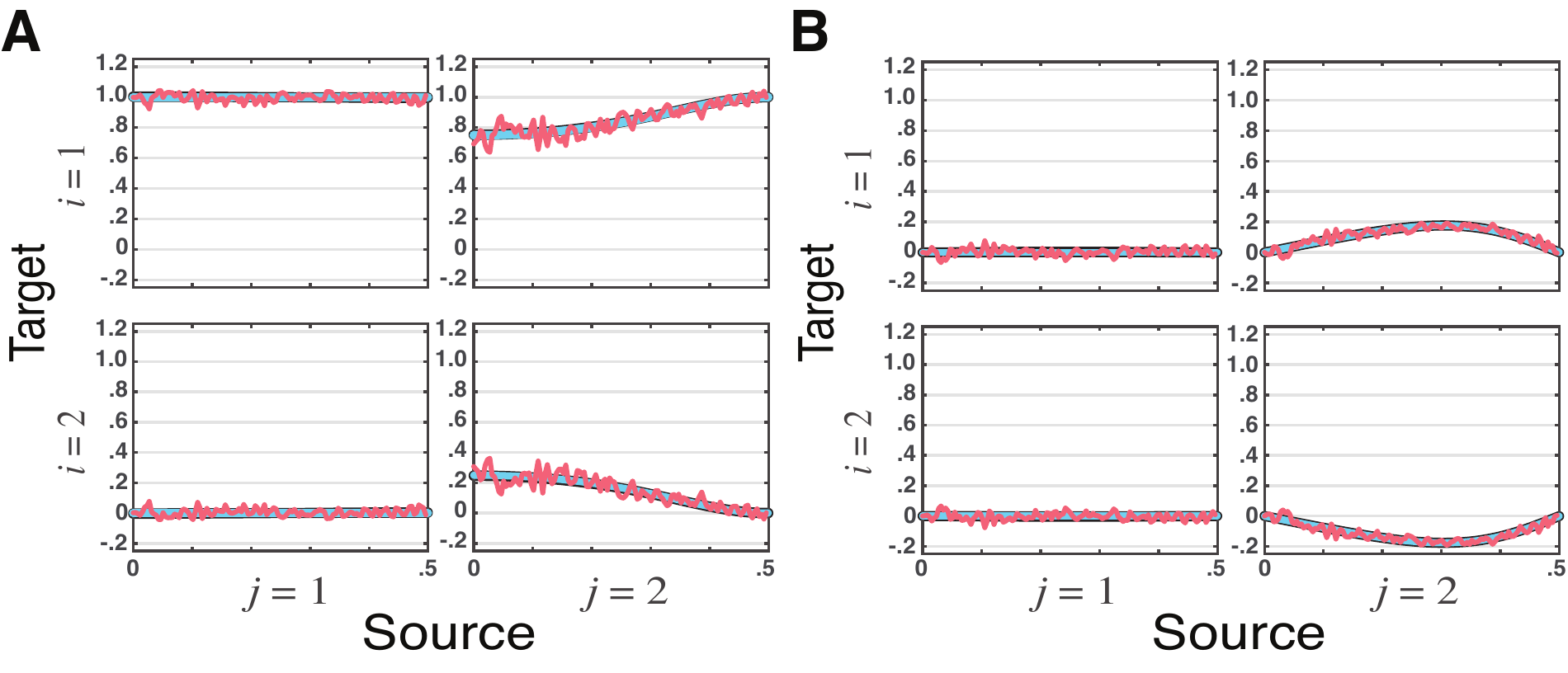} 
\end{center}
\caption{Superimposed graphs of \textbf{total partial directed coherence}, \textbf{tPDC},  estimates for the VMA model simulated for $n_s=16,384$ data points (Example \ref{ex:vma1}) and three  estimation methods (VAR, VMA, WN), where the \textbf{real} (\textbf{\textsf{A}}), and the \textbf{imaginary} (\textbf{\textsf{B}}) components are plotted separately. The theoretical tPDCs are depicted in \textcolor{Turquoise}{\textbf{blue lines}}.  WN estimates ripple around theoretical values (topmost \textcolor{RedOrange}{\textbf{red lines}}), yet they closely resemble that of theoretical values. VAR and VMA estimation methods results --- plotted as the two bottommost \textbf{black lines} --- are visually indistinguishable from the theoretical values (\textcolor{Turquoise}{\textbf{blue lines}}).}\label{fig:ex1}
\end{figure}
\end{ex}

\begin{ex}\textbf{Vector Autoregressive Moving Average Model (VARMA)}
\label{ex:varma}

The next example is a bit more elaborate. It has a VARMA(2,$\,$2) data generating procedure described by

\[ 
\left\{\begin{aligned}
x_1(n)&= 2\; r\; cos(\theta) x_1(n-1)-r^2x_1(n-2)+ w_1(n)+w_3(n) +w_3(n-1)\\\nonumber
x_2(n)&= b\; x_1(n-1)+ a \;x_2(n-1) + w_2(n)\\
x_3(n)&= c\; x_3(n-1)+ w_2(n)+ w_2(n-2)+w_3(n) \nonumber
\end{aligned} \right.
\label{eq:example2_varma} 
\]
where $r=.95$, $\theta=\pi/3$, $b=0.5$, $a=-.5$, $c=0.7$ and $\boldsymbol{\Sigma}_{\mathbf{w}}$ equal to the identity matrix.

As in the previous example,  total PDC estimates match one another regardless of method, see Figure \ref{fig:ex2}. 

Albeit at little surprise, it is important to realize that the use of the VARMA modelling scheme (Sec. \ref{sec:varmaMod}) yields substantially better fit. This is confirmed by  Table \ref{tab:errors} results.

\begin{figure}[h!]
\begin{center}
\includegraphics[width=.67\textwidth]{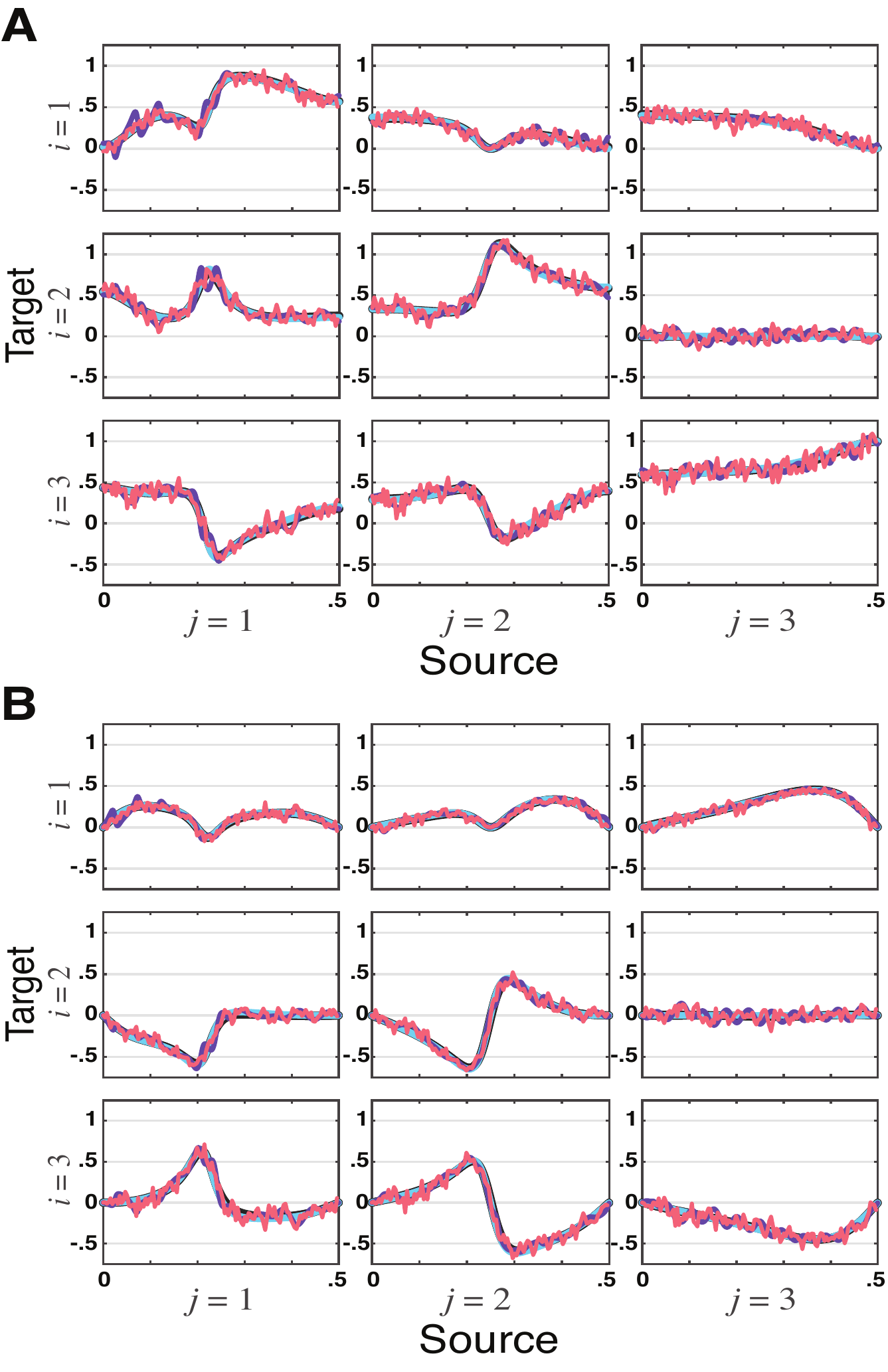}
\end{center}
\caption{~\textbf{tPDC} estimates by all four methods --- VAR, VMA, VARMA and WN --- for the VARMA model in Example \ref{ex:varma} simulated for $n_s=16,384$ data points are depicted,  with \textbf{real} (\textbf{\textsf{A}}), and \textbf{imaginary} (\textbf{\textsf{B}}) components plotted separately. As before, the theoretical tPDCs are also shown (\textcolor{Turquoise}{\textbf{blue lines}}). Again note that WN estimates (topmost \textcolor{RedOrange}{\textbf{red lines}}) ripple around theoretical values. In this case, VMA estimates (\textcolor{RedViolet}{\textbf{purple lines}}) also ripple around theoretical values (\textcolor{Turquoise}{\textbf{blue lines}}) illustrating estimator accuracy limitations. This is also apparent on Table \ref{tab:errors}. VAR and VARMA  results --- plotted as the two black bottommost lines just underneath the theoretical values --- represent much closer approximations.
}\label{fig:ex2}
\end{figure}
\end{ex}

\begin{ex} \textbf{Vector Autoregressive Model (VAR)}
\label{ex:var}

The third toy example covers the one used in \citep{entropy2021}  and was borrowed from \citep{faesLivro} involving three channels whose connectivity is assessed via a VAR model taking iGC effects into account through tPDC. One obtains essentially the same results irrespective of the computational approach,  see Figures \ref{fig:ex3}\textbf{\textsf{A}},\textbf{\textsf{B}}.  

\begin{figure}[h!]
\begin{center}
\includegraphics[width=.67\textwidth]{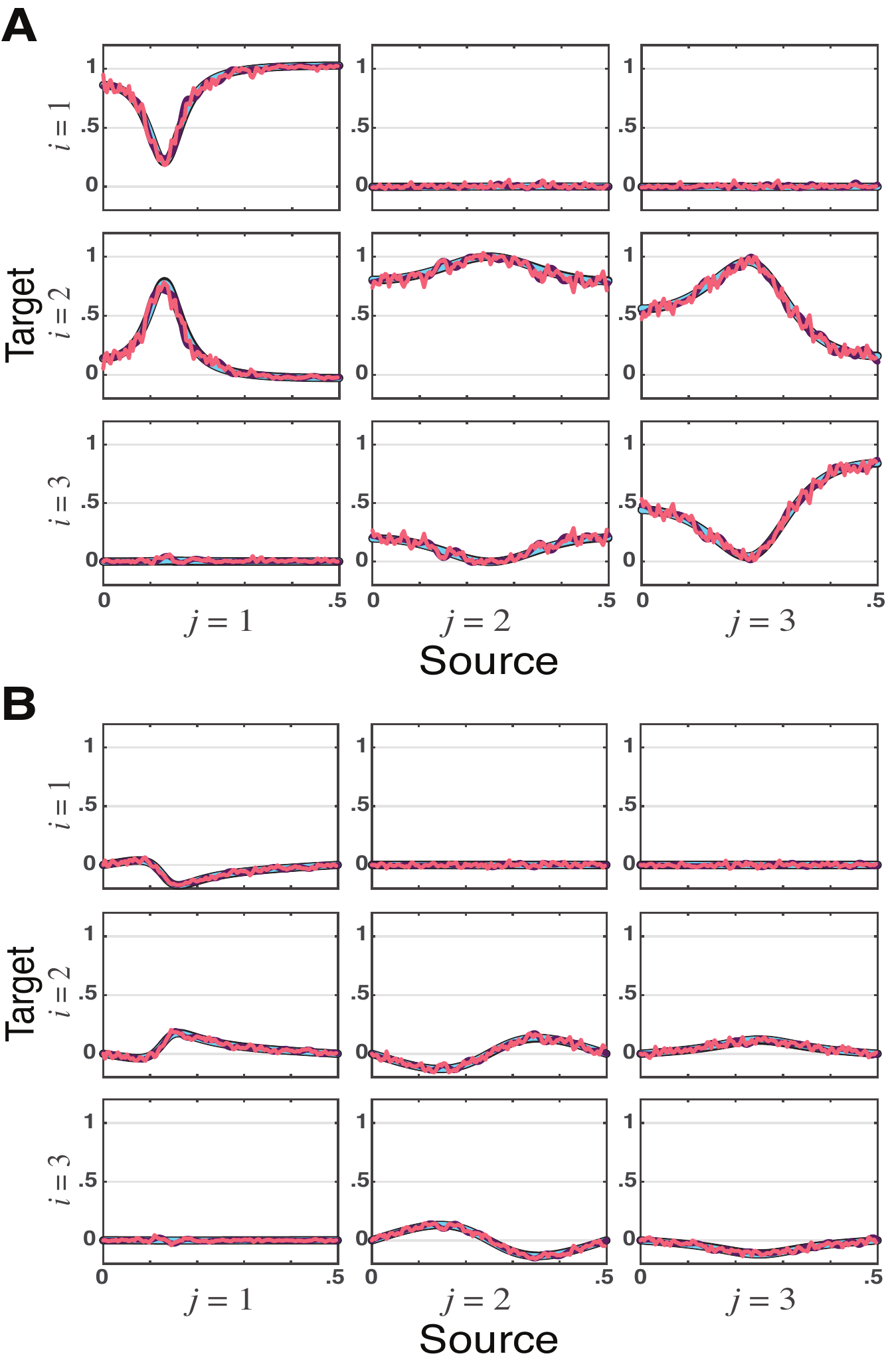}
\end{center}
\caption{\textbf{tPDC} estimates for Example \ref{ex:var} are shown for VAR, VMA and WN methods ($n_s=16,384$)  with \textbf{real} (\textbf{\textsf{A}}), and \textbf{imaginary} (\textbf{\textsf{B}}) components plotted separately. As before, theoretical tPDCs are also shown (\textcolor{Turquoise}{\textbf{blue lines}}). Once again, WN estimates (topmost \textcolor{RedOrange}{\textbf{red lines}}) ripple around theoretical values. Here so do too VMA estimates (\textbf{\textcolor{RedViolet}{purple lines}}) signalling their poor expected accuracy when fitting VAR data. This is confirmed by results presented on Table \ref{tab:errors}. VAR results are plotted as the two bottommost black lines underneath the theoretical values.}\label{fig:ex3}
\end{figure}

\end{ex}

\begin{ex}\textbf{Nonminimum Phase Data}
\label{ex:nonmim}

Consider a moving average data generation scheme using \eqref{eq:linmodel} with

\begin{equation}
\label{eq:nonmim_matrix}
\boldsymbol{\mathfrak{B}}_0 =\left[
\begin{array}{cc}
1 & 0\\
0 & 1
\end{array}
\right], \;
\boldsymbol{\mathfrak{B}}_1 =\left[
\begin{array}{cc}
2 & 1\\
0 & 0
\end{array}
\right], \;
\boldsymbol{\mathfrak{B}}_2 =\left[
\begin{array}{cc}
4 & 2\\
0 & 2
\end{array}
\right], \;
\end{equation}
whose allied \eqref{eq:detB} roots $\{-1+\pm \mathbf{j} \sqrt{3},\pm \mathbf{j} \sqrt{2} \}$  have magnitudes that are larger than 1, making this a nonminimum phase data generating mechanism as opposed to all previous examples, as computing their \eqref{eq:detB} easily shows. It is clear from \eqref{eq:nonmim_matrix} that $x_2(n)$ Granger-causes $x_1(n)$ but not otherwise. This is reflected in the computed tPDCT blue lines of Figure \ref{fig:ex4}. Here, \eqref{eq:vma1_pf} was adopted as the  innovations covariance matrix.

Use of Sec. \ref{sec:estim} algorithms leads to the results of Figure \ref{fig:ex4} where the estimation methods agree among themselves, but are markedly different from the tPDC computed using \eqref{eq:nonmim_matrix}. 

The reader may  easily verify using the Supplemental Material that the estimated solution using VMA modelling leads to \eqref{eq:detB} roots whose magnitudes are all smaller than 1.

Most importantly, however is that this example shows that GC causal relationships imposed through nonminimum phase systems  can be wrongly inferred. The consequences of this are further elaborated in the discussion.
\begin{figure}[h!]
\begin{center}
\includegraphics[width=.95\textwidth]{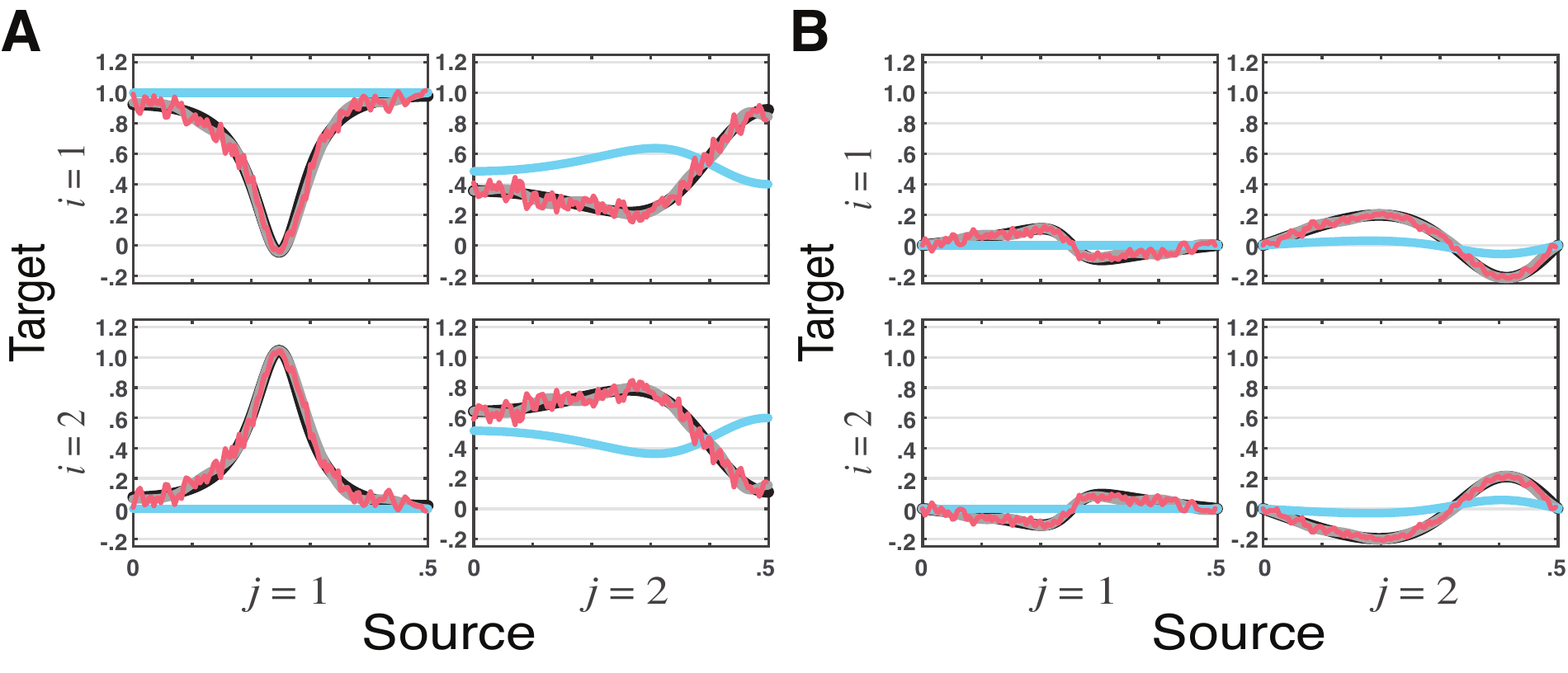}
\end{center}
\caption{\textbf{tPDC} estimates forVARMA model with nonminium phase data in  Example\ref{ex:nonmim} using the  VAR, VMA and WN methods ($n_s=16,384$)  portraying its (\textbf{\textsf{A}}) \textbf{real} and (\textbf{\textsf{B}}) \textbf{imaginary}  components. As before theoretical tPDCs are shown as \textcolor{Turquoise}{\textbf{blue lines}}. Here, WN estimates (topmost \textcolor{RedOrange}{\textbf{red lines}}) also ripple and agree with VMA (\textbf{\textcolor{black}{black lines}}) and VAR estimates (\textbf{\textcolor{darkgray}{gray lines}}) are very close to one another but differ significantly from the theoretical values (\textcolor{Turquoise}{\textbf{blue lines}}). Note the parameters in (\ref{eq:nonmim_matrix}) imply no connection from $x_1(n)$ onto $x_2(n)$, yet all three estimation methods wrongly indicate a non zero tPDC real component reflecting strong estimated GC.}\label{fig:ex4}
\end{figure}
\end{ex}

\begin{table}
\begin{center}
\begin{tabular}{crcccr} \toprule
Example & $n_s\quad$ & VMA & VAR & VARMA &  WN$\quad\ $\\
\hline\\
           & $16,384$ & $1.27 \times10^{-5}$ & $6.84 \times10^{-6} $& &  $0.15 \times10^{-2}$\\
 1            & $4,096 $& $5.76 \times10^{-5}$ & $3.06 \times10^{-5} $& &  $0.61  \times10^{-2}$\\
             & $1,024$ & $2.45 \times10^{-4}$ & $1.57 \times10^{-5} $& & $ 3.26  \times10^{-2}$\\
\midrule
          & $16,384$ & $0.21\times10^{-2}$ & $1.72 \times10^{-4} $& $2.96\times 10^{-8}$&  $0.67 \times10^{-2}$\\
2             & $4,096 $& $0.84 \times10^{-2}$ & $7.09 \times10^{-4} $& 
             $5.02\times 10^{-7}$ & $2.77  \times10^{-2}$\\
             & $1,024$ & $3.10 \times10^{-2}$ & $2.60 \times10^{-3} $& $6.65 \times10^{-6} $& $ 12.36 \times10^{-2}$\\
                                                 
\midrule
           & $16,384$ & $6.11\times10^{-4}$ & $3.20 \times10^{-5} $& & $0.13 \times 10^{-2}$\\
 3            & $4,096 $& $0.20 \times10^{-2}$ & $1.37 \times10^{-4} $&  & $0.57  \times10^{-2}$\\             
             & $1,024$ & $0.90 \times10^{-2}$ & $5.30 \times10^{-4} $& & $ 3.50  \times10^{-2}$\\ \bottomrule
                          
\end{tabular}
\end{center}
\caption{Table containing means squared error to fits of the theoretical tPDC according to estimation method for each Example. Missing values portray when certain estimation approaches were not used.}\label{tab:errors}
\end{table}

\section{Discussion}
\label{sec:discuss}

It is perhaps surprising that PDC/DTF have so long, and unnecessarily so, remained inextricably associated with VAR modelling even in view of early evidence to the contrary ~\citep{Jachan2009}. Partial explanation may lie in the early virtual exclusive reliance on VAR modelling that also dominated initial approaches to Granger Causality characterization \citep{Granger1969,Geweke1984}. This scenario in connection to time series modelling in the time domain slowly changed as VMA and VARMA approaches have been shown viable and possibly desirable depending on the nature of the data under study \citep{Boudjellaba-1992,Hafid-1994}. The latter methods are attractive because they more parsimoniously fit the underlying data as in Example \ref{ex:varma} via fewer parameters. This reflects Parzen's Parsimony Principle which formalizes the statistical advantage of describing data via the least possible number of parameters \citep{yaffee2000} that in the present case leads to lower average estimation error (see Table \ref{tab:errors}).  More details on alternative time domain characterization can be appreciated in\citet{Lutkepohl2005}.

Because of its prediction improvement ethos, Granger Causality, when originally defined,  rested on VAR modelling's predictive ability \citep{Granger1969}. Moreover, at that time it was the only practical alternative from a computational perspective. It is thus unsurprising that other predictive methods like VMA and VARMA modelling also can fit the purpose.

Given PDC/tPDC's frequency domain ties with Granger causality ~\citep{Baccala2021}v(with the inclusion of full instantaneous effects) \citep{entropy2021}, it is therefore no wonder that they too can be carried out via other methods like VMA or VARMA modelling. 

Thus we have shown that PDC/DTF (total or otherwise) are \textbf{not} irrevocably tied to VAR data modelling, though today, VAR remains the best studied and most widely applied option. It has the advantage of having rigorous asymptotic results in the squared PDC/DTF case \citep{Baccala2013,baccala_directed_2016}. Work is in progress to provide the asymptotics to the allied total PDC/DTF quantities.

Further research is needed to pinpoint which of the latter methods is best for what purpose. It is comforting to know that many methods provide equivalent descriptions if used properly. 

For example, even though it is possible to combine the response of different trials in event-related experiments while employing VAR models \citep{rodrigues_new_2015}, this feat may also, and perhaps more easily in some cases, be achieved through the application of Wilson's method to estimate nonparametric spectra and cross-spectra averaged over trials. Other methods have been proposed to deal with spectral matrix factorization that still need proper practical appraisal \citep{Amblard2015}. 

Though Wilson-type spectral factorization methods seem less effective in practice, it does not mean that they should be discarded. Here we only used Welch's spectral estimator. More research is needed, by employing other spectral estimation procedures like multitappering for instance \citep{percival_book} that could improve accuracy as they may more appropriately fit certain spectral shapes.

Here we have employed large data sets, but one should expect substantial performance differences for shorter time series. In this case, too, as hinted by Table \ref{tab:errors} results, VAR methods remain quite efficient, except when better approximation can be made through models that portray the data more closely as in the  VARMA (Example \ref{ex:varma}).

Other approaches have been proposed to obtain Granger-type estimates, namely state space modelling is one such example \citep{Barnett2015}; present research is on-going to evaluate them. In fact, as \citet{Sayedsurveyspectralfactorization2001}'s theoretical appraisal of univariate spectral density factorization methods suggests, even state-space models can be seen as spectral factorization providers.

All the above methods, by providing minimum phase spectral factors to the spectral density matrix, ideally portray \textbf{identical} Granger relationship representations within the accuracy and characteristic limitations of the employed spectral estimation/factorization techniques . 

However, there is an important caveat as we have shown. Due to their intrinsic minimum phase limitation,
the methods we explored here are unable to properly capture GC-type relations when the underlying data generation mechanism is nonminimum phase as in Example \ref{ex:nonmim}. This is due to the fact that these methods, either through classical time series modelling or direct spectral factorization, employ only second order statistics.

Though we do not show this explicitly here, Geweke-based approaches also suffer from the same limitations. This is easy to realize if one takes into account that they lead to conclusions that are similar to those reached via PDC/DTF-type approaches.

This scenario evokes two intertwined questions: (a) whether dynamical (viz. physical, physiological or economical) observations of phenomena actually conform to nonminimum phase generation mechanisms that might obscure their connectivity inference and (b) whether real data using GC methods in the past actually hold in view of this observation.

As an example consider a situation when nonminimum phase signals are a practical reality. It happens
in wireless communication, and is due to signal propagation through dispersive multipathway media that leads to serious bit-error rate impairment. As a man made system, this problem is circumvented by the transmission of pre-arranged pseudo random data (training) sequences the receiver uses to estimate channel nonminimality. Use of these sequences maps the receiver `output-only' problem into an equivalent `input-output' problem that can reveal nonminimum phase effects through second order statistic alone. This solution is sometimes unsatisfactory as  it imposes a penalty on the transmission rate of useful data. During the 1990's a considerable body of literature appeared to address this problem by dispensing with training sequences and using the received (output) data only ~\citep{haykin1994}. This is possible when the data is nongaussian, i.e. there is information beyond the ordinary second order statistics of the spectrum, something that can be made by design in telecommunication systems. Signal diversity in both time and space, via telecom signal characteristics or through employment of redundant receiver antennas is also an option. This general field has been known as that of  `\textit{blind}' identification/equalization \citep[see][for an overview]{chines2006}. Whereas real data properties cannot be made `designed' as in man made systems, they are often nongaussian and this could in principle be exploited to overcome the nonminimum phase generation limitation on GC inference we described here. 

The answer to (b) must thus await further analysis in what is a matter for further exciting research that may entail the revision of many conclusions regarding formerly analysed real data. 

\section{Conclusion}
\label{sec:conclu}

The first take home lesson is that PDC/DTF-type estimators of Granger connectivity/influentiability \citep{baccala_causality_2014} even in their latest and most general total form (tPDC/tDTF), incorporating instantaneous Granger effects, do not require vector autoregressive modelling as a mandatory step but can be obtained through any other means of spectral factorization of the spectral density matrix into minimum phase factors. The second lesson is that, though not mandatory, VAR modelling, since it can be used to obtain consistent spectral factors, and because of its practicality and efficiency, remains the method of choice, specially for short data sets. The third no less important lesson is that care as to conclusions about real data must be exercised as possible unknown nonminimum phase data generating mechanisms may be at play that can confound results as to the actual true underlying connectivity when methods of the present spectral factorization class are used.

\section*{Supplemental Material}

As supplemental material, the MATLAB codes used to generate the four figures in this manuscript are made available [\href{https://www.dropbox.com/s/psphg7pgc7tiic1/PDCVARMYTH2022.zip?dl=0}{here}]. To begin with, run startup.m that will set up the paths and verify for the presence of required MATLAB toolboxes, then you may execute Example1.m, Example2.m, Example3.m and Example4 scripts from the command line to get the respective figures. These routines are distributed under GNU General Public License v3.0.

\section*{Conflict of Interest Statement}

The authors declare that the research was conducted in the absence of any commercial or financial relationships that could be construed as a potential conflict of interest.

\section*{Author Contributions}

Both authors have equally shared in the conception, writing and editing the paper.

\section*{Funding}
L.A.B. was funded by CNPq, grant number 308073/2017-7. L.A.B and K.S. were partially supported by FAPESP Grant 2017/12943-8.

\section*{Acknowledgments}
K.S. is affiliated with LIM 43–HCFMUSP. Both are attached to the Center for Interdisciplinary Research on Applied Neurosciences (NAPNA), Universidade de São Paulo, São Paulo, Brazil.
.



\bibliographystyle{plainnat}  
\bibliography{references}

\end{document}